\pdfoutput=1
\documentclass[twocolumn,prd,floatfix,preprintnumbers,nofootinbib]{revtex4}
 
\usepackage{graphicx}
\usepackage{bm}
\usepackage{color}
\usepackage{amsmath}
\usepackage[hidelinks]{hyperref}
\usepackage{doi}

\begin{document}

\title{Fast neutrino conversions: Ubiquitous in compact binary merger remnants}

\author{Meng-Ru Wu}
\email{wu@nbi.ku.dk}
\affiliation{Niels Bohr International Academy, Niels Bohr Institute, Blegdamsvej 17, 2100 Copenhagen, Denmark} 
\author{Irene Tamborra}
\email{tamborra@nbi.ku.dk}
\affiliation{Niels Bohr International Academy, Niels Bohr Institute, Blegdamsvej 17, 2100 Copenhagen, Denmark} 

\date{\today}

\begin{abstract}
The massive neutron star (NS) or black hole 
(BH) accretion disk  resulting from NS--NS or 
NS--BH mergers is dense in neutrinos. We present the 
first study on the role of angular distributions in 
the neutrino flavor conversion above the remnant disk. 
In particular, we focus on ``fast'' pairwise conversions 
whose rate depends on the local angular intensity of the 
electron lepton number carried by neutrinos.  
Because of the emission geometry and the  
flux density of $\bar{\nu}_e$ being larger than that of $\nu_e$, 
fast conversions prove to be a generic phenomenon in NS--NS 
and NS--BH mergers for physically motivated disturbances 
in the mean field of flavor coherence. 
Our findings suggest that, differently from the core-collapse 
supernova case, fast flavor conversions
seem to be unavoidable in compact mergers and could 
have major consequences for the jet dynamics and the 
synthesis of elements above the remnant disk.
\end{abstract}

%\pacs{14.60.Pq, 97.60.Bw}
%\keywords{}

\maketitle

\section{Introduction}\label{sec-intro}
The mergers of one neutron star (NS) 
with another NS or a black hole (BH) are among the most 
promising astrophysical sites to account for the observed 
short gamma-ray bursts (sGRB)~\cite{Eichler:1989ve,Berger:2013jza}, 
the potential kilonova/macronova 
candidates~\cite{Metzger:2010sy,Tanvir:2013pia,Yang:2015pha}, 
and the production of heavy elements above iron
in our Universe~\cite{1974ApJ...192L.145L,Eichler:1989ve,1999ApJ...525L.121F}. 
They are also expected to be sources of gravitational 
waves in addition to the BH--BH mergers recently detected~\cite{Abbott:2016blz,Abbott:2016nmj}.

Similarly to core-collapse supernovae (SNe), 
the merger remnant accretion disks surrounding the central massive NS or BH, 
are neutrino-dense sites. 
A vast number of neutrinos is produced in this hot and dense
environment formed during the dynamical (post)merging phase. 
The total neutrino energy luminosity can reach
$\sim 10^{53}$--$10^{54}$~erg/s at peak for
$\sim 100$~ms~\cite{Ruffert:1996by,Foucart:2015vpa,Perego:2014fma}.

Despite the fact that the estimated merger rate within the 
detection volume of current and upcoming large-scale neutrino detectors 
is low~\cite{Sadowski:2007dz,Abbott:2016ymx}, neutrinos play
an important role in various physical processes happening
during and after the merging. 
For instance, the absorption of neutrinos on the matter ejected 
dynamically during the NS--NS merger phase and the 
postmerger massive NS accretion
disk may largely affect the neutron richness of 
the ejecta~\cite{Wanajo:2014wha,Metzger:2014ila,Perego:2014fma}. 
Consequently, neutrinos may alter the nucleosynthesis outcome of the
rapid-neutron capture process ($r$-process)
and the associated kilonova light curves.
Moreover, the question of whether the pair annihilation of neutrinos 
above the BH accretion disk resulting from the mergers
can be a viable option for the sGRB jet formation remains 
unsettled~\cite{Birkl:2006mu,Zalamea:2010ax,Richers:2015lma,Just:2015dba}.

Given the potential major role of neutrinos in merger remnants, 
understanding their flavor evolution is crucial. 
Besides ordinary interactions of neutrinos with 
matter~\cite{Wolfenstein,Mikheyev:1986gs}, 
the $\nu$--$\nu$ coherent forward 
scattering~\cite{Fuller:1987,Pantaleone:1992eq,Sigl:1992fn} can affect 
the flavor composition. However, while neutrino oscillations in the 
SN context have been widely discussed in the 
literature~\cite{Duan:2010bg,Duan:2015cqa,Mirizzi:2015eza,Chakraborty:2016yeg}, 
only preliminary work exists on flavor conversions above 
mergers~\cite{Malkus:2012ts,Malkus:2014iqa,Malkus:2015mda,Zhu:2016mwa,Frensel:2016fge,Chatelain:2016xva}. 

In mergers, a phenomenon known as ``matter-neutrino resonance'' 
(MNR)~\cite{Malkus:2014iqa,Malkus:2015mda,Wu:2015fga} 
may occur due to the near cancellation of the large but opposite contributions
of $\nu$--e and $\nu$--$\nu$ interaction energies. The latter being 
negative as $\propto n_{\nu_e}-n_{\bar{\nu}_e} <0$, with $n_{\nu_e}$ 
($n_{\bar{\nu}_e}$) the local $\nu_e$ ($\bar{\nu}_e$) 
number densities; such a condition results from the overall protonization 
of the merger remnants~\cite{Ruffert:1996by,Foucart:2015vpa,Perego:2014fma}.

The leading role played by the neutrino angular 
distribution in $\nu$--$\nu$ interactions has been
appreciated only recently in SNe~\cite{Duan:2010bg,Mirizzi:2015eza,Chakraborty:2016yeg}. 
In particular, close to the neutrino decoupling region in SNe, ``fast'' neutrino 
conversions~\cite{Sawyer:2005jk, Sawyer:2008zs, Sawyer:2015dsa}
may occur within the length of 
$\sim(G_F |n_{\nu_e}-n_{\bar{\nu}_e}|)^{-1} \simeq \mathcal{O}(10)$~cm,
with $G_F$ the Fermi constant. 
Fast conversions may quickly lead to flavor equilibration and 
are exclusively driven by the angular distribution of
the electron neutrino lepton number (ELN). 
Nevertheless, the angular distribution has 
been integrated out in all existing oscillation studies in merger remnants. 

Similarly to SNe, the $\bar\nu_e$ decoupling region resides 
inside the $\nu_e$ one in merger remnants.
However, differently from SNe, the flux of ${\bar{\nu}_e}$ is larger than 
that of ${\nu_e}$ because of the overall protonization. 
These lead to crossings between the angular distributions 
$\Phi_{\bar\nu_e}=d n_{\bar\nu_e}/d\Omega$ and $\Phi_{\nu_e}=d n_{\nu_e}/d\Omega$ 
(i.e., changes of sign of the ELN distribution 
$\Phi_{\bar\nu_e}-\Phi_{\nu_e}$)
at any point above the $\nu_e$ decoupling surface 
as illustrated in Fig.~\ref{fig-gv1}.
Fast neutrino conversions due to temporal instabilities should, therefore, be 
expected~\cite{Chakraborty:2016lct,Dasgupta:2016dbv,Izaguirre:2016gsx}. 
Note that in SNe,  crossings of ELN distribution 
are not guaranteed (see e.g.~Ref.~\cite{Tamborra:2017ubu}); they may only occur in the 
presence of LESA for certain emission 
directions~\cite{Izaguirre:2016gsx,Tamborra:2014aua}. 
Therefore, fast conversions in supernovae
may mainly occur because of the non-negligible flux of neutrinos not streaming in the radially forward 
direction~\cite{Izaguirre:2016gsx}.
In this sense, the merger remnants offer a more natural environment 
than SNe for fast conversions.

In this paper, the neutrino angular distributions are taken into account 
in the study of $\nu$--$\nu$ interactions above merger remnant disks for 
the first time. Similarly to core-collapse SNe, an exact numerical solution 
of the flavor distribution of propagating neutrinos 
in mergers is not yet affordable. 
However, we can estimate whether favorable conditions for fast flavor conversions
are present in compact binary merger remnants by adopting analytical tools. To this purpose, we rely on the dispersion relation (DR) approach
recently developed in Ref.~\cite{Izaguirre:2016gsx}.

The outline of our manuscript is as follows. First, we model the neutrino emission from compact binary
merger remnants by introducing a simple two-neutrino-emitting 
disk model motivated by existing hydrodynamical simulations in Sec.~\ref{sec:disk}.
In Sec.~\ref{sec:EoM}, we introduce the equation of motion governing the neutrino flavor evolution 
and the DR in the flavor space. Results on the occurrence of temporal and 
spatial instabilities in the flavor space are presented in Sec.~\ref{sec:temporal} and Sec.~\ref{sec:spatial}, respectively. 
Caveats on our main findings are discussed in Sec.~\ref{sec:discussion}, and conclusions are reported in Sec.~\ref{sec:conclusions}.

\section{Two-neutrino-emitting disk model}\label{sec:disk}
\begin{figure}[t]
 \includegraphics[angle=0,width=1.\columnwidth]{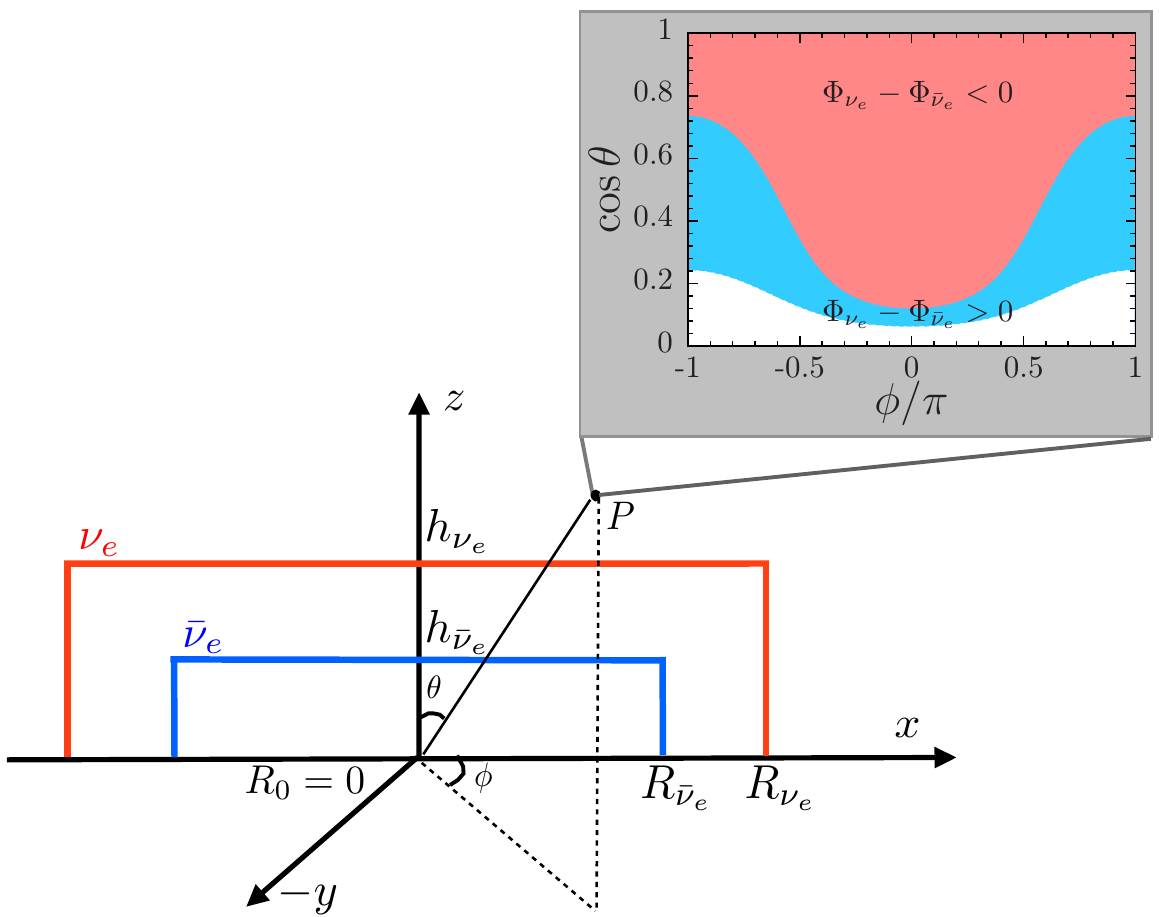}  
  \caption{Geometry of $\nu_e$ (in red) and $\bar\nu_e$ (in blue) emitting
  surfaces with radii $R_{\nu_e}$ and $R_{\bar\nu_e}$,
  heights $h_{\nu_e}$ and $h_{\bar\nu_e}$.
  $R_0$ is the innermost stable circular orbit for a BH-disk system ($R_0=0$ 
  for a NS-disk remnant). Inset: Example of crossings of the ELN distribution
  ($\Phi_{\nu_e}- \Phi_{\bar\nu_e}$) as a 
  function of the polar and azimuthal angles $\cos \theta$ and 
  $\phi$ above the NS--disk. The exact shapes are calculated at  
  $(x,z)=(0.6 R_{\nu_e},0.35 R_{\nu_e})$ with
  $R_{\bar\nu_e}=0.75~R_{\nu_e}$ and 
  $h_{\nu_e}/R_{\nu_e}=h_{\bar\nu_e}/R_{\bar\nu_e}=0.25$. 
  The region shaded in red 
  (blue) corresponds to $\Phi_{\nu_e} < \Phi_{\bar\nu_e}$ 
  ($\Phi_{\nu_e} > \Phi_{\bar\nu_e}$).  
  \label{fig-gv1}}
\end{figure}

In order to examine whether fast flavor conversion 
occurs above the merger remnants, 
we refrain from relying on a specific merger model 
given the uncertainties intrinsic to the neutrino transport 
adopted in hydrodynamical simulations of these objects. 
We instead rely on the simple two-neutrino-emitting 
disk model shown in Fig.~\ref{fig-gv1} (see also Appendix A). 
The choice of the model parameters 
is, however, guided by the hydrodynamical simulation of the 
massive NS--disk evolution~\cite{Perego:2014fma}. 

In addition to the overall protonization discussed in the previous section, 
an important feature of merger remnants is that the 
spectral-averaged decoupling surfaces of $\nu_e$ and $\bar\nu_e$ are spatially
well separated. This can be seen, for example, in
Fig.~12 of Ref.~\citep{Foucart:2015vpa} and Fig.~3 of Ref.~\citep{Frensel:2016fge}
showing the size ratio of the decoupling surface of $\bar\nu_e$ to 
that of $\nu_e$ $\sim 3/4$.
This is a consequence of the neutron richness of 
the remnant system and the spatial extension of the accretion 
disk which leads to a smaller density gradient 
with respect to the SN proto-neutron star.

Based on the above discussion, we assume that for a NS--disk remnant, 
$\nu_e$ and $\bar\nu_e$ decouple instantaneously at 
surfaces approximated as finite-size disks of radii
$R_{\bar\nu_e}=0.75 R_{\nu_e}$ and heights 
$h_{\nu_e}/R_{\nu_e}=h_{\bar\nu_e}/R_{\bar\nu_e}=0.25$. 
They are emitted half-isotropically from their respective 
surfaces with a flux ratio 
$\alpha\equiv\Phi^0_{\bar\nu_e}/\Phi^0_{\nu_e}=2.4$ and
propagate freely afterwards. 
For the BH--torus, we model the 
$\nu$-emitting tori by setting an inner edge
of the surface at $R_0=0.15 R_{\nu_e}$~\cite{Foucart:2015vpa}, 
representing the innermost stable circular orbit. 
Since in the merger remnants, the nonelectron neutrinos 
share the same properties, they
do not enter the following analysis and will
be omitted.

\section{Dispersion relation in flavor space} \label{sec:EoM}
The equation of motion (EoM) for each momentum mode 
governing the evolution of free streaming neutrinos is given by:
$(\partial_t+\mathbf{v}\cdot\partial_{\mathbf x})\varrho
=-i[H,\varrho]$
where $\mathbf{v}=(\sin\theta\cos\phi,\sin\theta\sin\phi,\cos\theta)$ 
is the velocity of an ultrarelativistic neutrino, 
whose 4-vector is $v^\mu=(1,\mathbf{v})$. 
The Wigner-transformed density matrix $\varrho$ 
in the flavor basis encodes the flavor occupation numbers 
in the diagonal terms and flavor correlations in the off-diagonal terms. 
The Hamiltonian, $H$, consists of the contributions 
from the vacuum mixing~\cite{Olive:2016xmw}, 
coherent-forward scattering between neutrinos and electrons,
and that among neutrino themselves.

Dismissing the vacuum term and ignoring the energy dependence 
since we are interested in fast conversions, 
we express the neutrino density matrix in terms of the 
``flavor isospin'' $\xi$ and 
the occupation numbers $f_{\nu_\beta}$ for the neutrino
flavor $\nu_\beta$:
$\varrho=[(f_{\nu_e}+f_{\nu_x})+(f_{\nu_e}-f_{\nu_x})\xi]/2$ 
($\bar\varrho=-[(f_{\nu_e}+f_{\bar\nu_x})+
(f_{\bar\nu_e}-f_{\bar\nu_x})\xi^*]/2$)
for neutrinos (antineutrinos)~\footnote{Quantities such as $f_{\nu_\beta}$, 
$n_{\nu_\beta}$ and $\Phi_{\nu_\beta}$ are defined in 
the absence of flavor conversions in this work.}
under the two-flavor mixing approximation.
The Hamiltonian for $\xi(\mathbf{v})$ can be written as
\begin{equation}\label{eq-eomnfis}
H = v^\mu \lambda_\mu \frac{\sigma_3}{2}+ \int d\Omega^\prime 
v^\mu v^\prime_\mu \xi(\mathbf{v}^\prime)g(\mathbf{v}^\prime)\ ,
\end{equation}
where, given the metric $\eta^{\mu\nu}={\rm diag}(1,-1,-1,-1)$, 
$v^\mu \lambda_\mu = \lambda_0 - \mathbf{v} \cdot \bm{\lambda}$ 
with $\lambda_0 = \sqrt{2} G_F n_e$, 
$n_e$ being the net electron number density,
$\bm\lambda=\lambda_0\mathbf{v_f}$, and
$\mathbf{v_f}$ being the local fluid velocity.
The neutrino potential angular distribution $g(\mathbf{v})$ 
per unit length per unit solid angle is proportional to %is defined by means of 
the ELN angular distribution
\begin{equation}
\label{eq:disk}
g(\mathbf{v})\!=\! \sqrt{2}G_F
\left(\Phi_{\nu_e} - \Phi_{\bar{\nu}_e}\right)\ .
\end{equation}
$\Phi_{\nu_\beta}$ is related to the distribution functions $f_{\nu_\beta}$ by
\begin{equation}
\Phi_{\nu_\beta}(\mathbf{v}) = 
\frac{dn_{\nu_\beta}(\mathbf{v})}{d\Omega}\!
=\! \frac{1}{(2\pi)^3} \int dE E^2 f_{\nu_\beta}(E,\mathbf{v})\ ,
\end{equation}
with $d\Omega=d\cos\theta d\phi$  the differential solid angle.
Since we work in the corotating frame of the disk, 
the term $\mathbf{v} \cdot \bm\lambda$ 
will be neglected from now on.  

In order to investigate whether off-diagonal terms may originate 
in the density matrix giving rise to fast conversions, we now linearize 
the EoM~\cite{Banerjee:2011fj,Raffelt:2013rqa} and
track the evolution of the off-diagonal term $S$ in $\xi$,
\begin{equation}
\xi=\left(\begin{matrix}
1 & S\\
S^* & -1\\
\end{matrix}
\right),
\end{equation}
by neglecting terms larger than $\mathcal{O}(|S|)$. 
Assuming that $S(\mathbf{v})$ evolves as a plane wave 
$S(\mathbf{v},t,{\mathbf x})=Q(\mathbf{v},\omega,\mathbf k)
e^{-i(\omega t- {\mathbf k}\cdot{\mathbf x})}$, 
the EoM becomes~\cite{Izaguirre:2016gsx}
\begin{equation}\label{eq-eom2}
v_\mu s^\mu Q(\mathbf{v},\omega,\mathbf k)
+\int d\Omega^\prime
v_\mu v^{\prime\mu}g(\mathbf{v}^\prime)
Q(\mathbf{v^\prime},\omega,\mathbf k)=0\ ,
\end{equation}
by defining the 4-vector 
$s^\mu\equiv (\omega-\lambda_0-\epsilon_0,
\mathbf{k}-\bm{\epsilon})$,
$\epsilon_0\equiv \int d\Omega g(\mathbf{v})$ and 
$\bm{\epsilon} \equiv \int d\Omega g(\mathbf{v})\mathbf{v}$. 
From the structure of Eq.~\eqref{eq-eom2}, one sees that the solution of 
$Q(\mathbf{v},\omega,\mathbf k)$ has the form
of $(v_\mu a^\mu)/(v_\mu s^\mu)$.
A nontrivial $a^\mu$ exists, if
\begin{equation}\label{eq-DR-general}
{\rm det}[\Pi_{\mu\nu}(\omega,\mathbf k)]=0\ ,
\end{equation}
where $\Pi_{\mu\nu}(\omega,\mathbf k)=\eta_{\mu\nu}+\int d\Omega\ v_\mu v_\nu 
g(\mathbf{v})/(v_\mu s^\mu)$.

Equation~\eqref{eq-DR-general} is the DR for
the mode with $(\omega,\mathbf k)$~\cite{Izaguirre:2016gsx}.
If the solutions satisfying the DR consist
of real $(\omega,\mathbf k)$ only, any initial perturbations
in the flavor space do not grow in the linear regime.
If, however, any conjugate pair of complex solutions 
in $\omega$ or $k_i$ ($i\in\{x,y,z\}$) satisfies the DR,  
an instability growing exponentially with  rate
$|{\rm Im}(\omega)|$ or $|{\rm Im}(k_i)|$
occurs, leading to  flavor conversions.
Practically, one examines whether the system
is temporally (spatially) unstable by finding all solutions
of $\omega$ ($k_i$) with a given $\mathbf{k}$ ($\omega$ and $k_{j\neq i}$)~\cite{Izaguirre:2016gsx}.

\begin{figure}[t]
  \includegraphics[angle=0,width=.9\columnwidth]{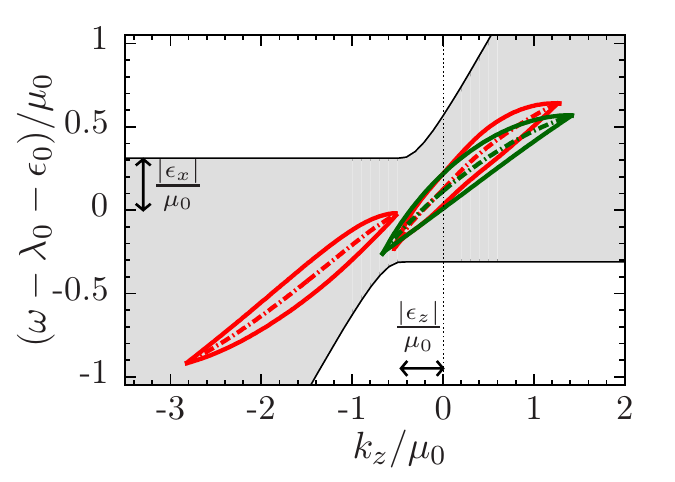}     
  \caption{Dispersion relation of $\mathbf{k}=(0,0,k_z)$ 
  for the ELN distribution 
  shown in Fig.~\ref{fig-gv1} with 
  $\alpha=\Phi^0_{\bar\nu_e}/\Phi^0_{\nu_e}=2.4$. 
  For the complex $\omega$ that lead to flavor instability, 
  ${\rm Re}(\omega)$ are shown by the red (green) dash-dotted curves
  and ${\rm Re}(\omega)\pm{\rm Im}(\omega)$ are shown by
  the red (green) solid curves for the MS
  preserving (breaking) solutions. The gray regions are the zone of 
  avoidance for real $(\omega,k_z)$. 
  The temporal instability exists for a large range of $k_z/\mu_0$ and it is, therefore, unavoidable.
  \label{fig-wkz}}
\end{figure}

\section{Temporal instabilities above the remnants}  \label{sec:temporal}
Motivated by the crossings of the local ELN 
distribution resulting from the disk emission geometry and the 
protonization of the merger remnant, 
we now investigate conditions for 
temporal instabilities above the NS--disk system. 
To this purpose, we examine the solution of the DR 
for the ELN distribution at 
the location $(x,z)=(0.6~R_{\nu_e},0.35~R_{\nu_e})$ 
above the NS--disk system shown in Fig.~\ref{fig-gv1}
as a benchmark case. 
We note that any generic point above the $\nu_e$-emitting surface
would exhibit qualitatively similar solutions for the DR 
(see also Appendix A for  
$g(\mathbf v)$ distributions at different locations),
despite the relative contribution of $\nu_e$ and $\bar\nu_e$ to
$g(\mathbf v)$ varies significantly at different locations.
This is because the features responsible for inducing flavor 
conversions (i.e., the change of sign in $g(\mathbf v)$) are 
present everywhere above the remnant  and cannot be
avoided differently from the SN case. 

First, we solve the DR equation for the modes $\mathbf{k}=(0,0,k_z)$
and $\alpha\equiv\Phi^0_{\bar\nu_e}/\Phi^0_{\nu_e}=2.4$.
Due to the mirror symmetry (MS) with respect to the $y$ axis, 
Eq.~\eqref{eq-DR-general} has $\Pi_{02}=\Pi_{12}=\Pi_{32}=0$ 
and it reduces to
\begin{subequations}\label{eq-DR}\label{sublab3}
\begin{align}
\Pi_{22}(-\Pi_{03}^2\Pi_{11}+2\Pi_{01}\Pi_{03}\Pi_{13}-
\Pi_{00}\Pi_{13}^2-\Pi_{33}\Pi_{01}^2\nonumber\\
+\Pi_{00}\Pi_{11}\Pi_{33})=0.\tag{\ref{sublab3}}
\end{align}
\end{subequations}
An ${\rm Im}(\omega)\neq 0$ from $\Pi_{22}=0$ corresponds to
a MS breaking unstable solution. When the terms inside the parentheses sum to 0, 
a MS-preserving 
solution occurs instead.

Figure~\ref{fig-wkz} shows the temporal DR for 
the benchmark location.
We define $\mu_0=\sqrt{2}G_F n^0_{\nu_e}$
to denote a typical length scale,
\begin{equation}\label{eq-mu0}
\mu_0\approx 4.25\ {\rm cm}^{-1}
\left(\frac{L_{\nu_e}}{10^{53}{\rm erg/s}}\right)
\left(\frac{10{\rm MeV}}{\langle E_{\nu_e}\rangle}\right)
\left[\frac{(100{\rm km})^2}{R_{\nu_e}^2-R_0^2}\right]\ ,
\end{equation}
with $L_{\nu_e}$ the total energy luminosity of $\nu_e$, 
$\langle E_{\nu_e}\rangle$ the corresponding average energy. 
A temporal instability generally exists for a 
large range of $|k_z|/\mu_0$ (i.e., $|k_z| \simeq$ a few times $\mu_0$),
centered around $k_z \approx \epsilon_z=-0.46~\mu_0$.

Because of the broken axial symmetry, 
we find three branches of the unstable solutions: 
two of them preserving the $y$-MS (red curves) and the other one 
breaking this symmetry (green curves). 
No real solutions exist for the selected $(x,z)$ 
due to the enlarged zone of avoidance (gray region), where $Q(\mathbf{v})$ would 
become singular as a result of larger asymmetry in the $x$ direction 
($|\epsilon_x|=0.31\mu_0$).
Note that, for $g(\mathbf{v})$ corresponding to other $(x,z)$ above the 
$\nu_e$ surface, we find nonzero real solution(s) that sometimes coexist 
with the complex ones because of the disk geometry, 
differently from the findings of Ref.~\cite{Izaguirre:2016gsx}.

\begin{figure}[t]
  \includegraphics[angle=0,width=1.\columnwidth]{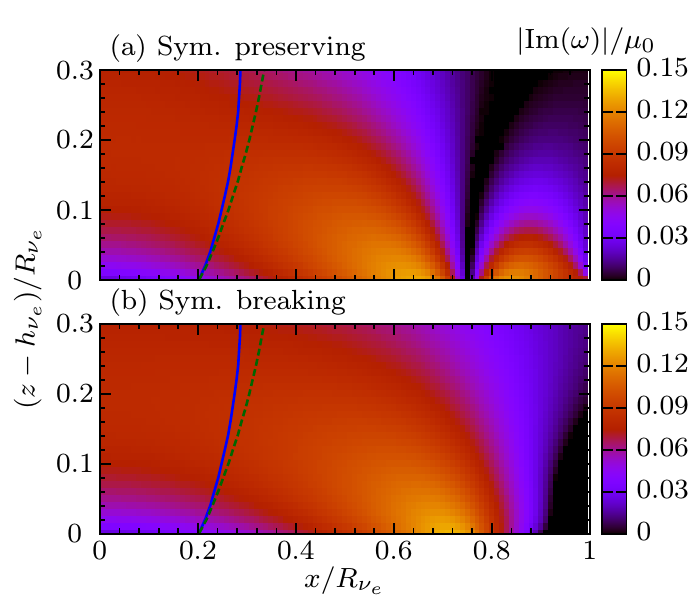}     
  \caption{Contour plot of $|{\rm Im}(\omega)|/\mu_0$ in the $(x,z)$ 
  plane above the $\nu_e$-surface for $\mathbf{k}=0$ for our benchmark
  NS-disk model. 
  The mirror symmetry preserving and breaking solutions
  are shown in panel (a) and (b).
  Also shown are the MNR locations obtained by adopting the $n_e$ profiles 
  of~\cite{Perego:2014fma} for neutrinos emitted at $(x,z)=(0,h_{\nu_e})$
  (solid blue curve) and at $(x,z)=(-R_{\nu_e},h_{\nu_e})$
  (dashed green curve). Fast conversions occur everywhere above the 
  $\nu_e$ surface and could alter the otherwise favorable conditions for the MNR.
  \label{fig-insdisk}}
\end{figure}

So far, we have focused on the temporal instabilities 
for one specific point $(x,z)$ above the $\nu_e$ surface of
the NS disk. Since the $k_z=0$ mode is unstable in this studied case, 
we adopt it as the benchmark wave number and 
generalize our study to any $(x,z)$ above the $\nu_e$ emitting disk surface. 
Figure~\ref{fig-insdisk} shows contour plots of the growth 
rate $|\mathrm{Im}(\omega_0)|/\mu_0$ in the $(x,z)$ plane. 
MS preserving (panel(a)) and breaking (panel(b)) flavor instabilities exist 
in most of the region above the $\nu_e$ surface.
Moreover, $|\mathrm{Im}(\omega_0)|/\mu_0$ at some vertical distance
above the $\nu_e$ surface may be larger than that at the surface.
 The growth rate $|{\rm Im}(\omega)|$ is typically 
large $\sim 0.1\mu_0$, leading to flavor conversion 
in the length scale of cm.

The eventual occurrence of fast conversions above the $\nu_e$ emitting 
surface hints towards a change of paradigm of the current 
picture of flavor conversions in merger remnants, in particular 
concerning the occurrence of the MNR. 
In fact, in Fig.~\ref{fig-insdisk} we also show the 
locations of the MNR for neutrinos emitted from the
center ($x=0, z=h_{\nu_e}$, solid blue curve) and the opposite 
side of the disk ($x=-R_{\nu_e}, z=h_{\nu_e}$, dashed green curve), 
using the $n_e$ profiles from~\citep{Perego:2014fma}
to model the matter potential.
Before these neutrinos reach the MNR locations,
they will traverse the region where the
fast conversions can occur and this would likely alter the MNR 
conditions. 

\begin{figure}[t]
  \center
  \includegraphics[angle=0,width=0.85\columnwidth]{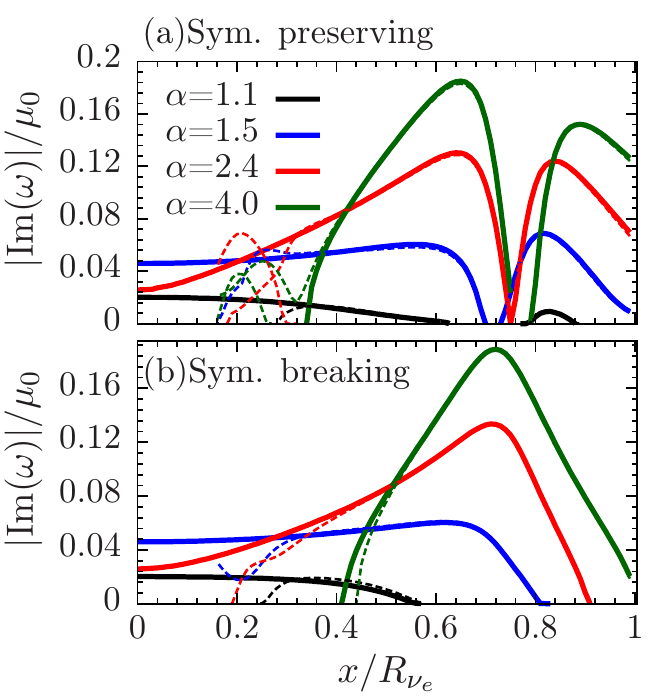}     
  \caption{Growth rate of the flavor instability, $|{\rm Im}(\omega)|/\mu_0$,
  as a function of $x$ on the $\nu_e$-surface for the same
  NS disk in Fig.~\ref{fig-gv1} (thick curves) with 
  different $\alpha=\Phi^0_{\bar\nu_e}/\Phi^0_{\nu_e}$
  for the mirror symmetry preserving solutions [panel (a)] and the mirror 
  symmetry breaking solutions [panel (b)]. The thin dashed curves show the 
  corresponding solutions of the BH torus with $R_0=0.15 R_{\nu_e}$. For 
  any $\alpha\geq 1.1$, flavor instabilities exist above the $\nu_e$ surface.
  \label{fig-insz0}}
\end{figure}

In order to further generalize our results to any $\alpha$ compatible 
with existing hydrodynamical simulations, Fig.~\ref{fig-insz0} 
shows the temporal instability of $\mathbf{k}=0$
for the same disk model with different values of 
the flux ratio $\alpha=1.1$, 1.5, 2.4, and 4.0
\footnote{We note that $\alpha>1.78$ ($\alpha<1.78$) corresponds to 
a net protonization (neutronization)
of the NS--disk system for the assumed $R_{\bar\nu_e}/R_{\nu_e}$.}
covering the range of 
$(L_{\bar\nu_e}/\langle E_{\bar\nu_e}\rangle)/
(L_{\nu_e}/\langle E_{\nu_e}\rangle)$
from hydrodynamical simulations 
listed in Table~7 of Ref.~\cite{Frensel:2016fge}.
For this wide range of $\alpha$,
MS preserving and breaking flavor instabilities are found in most of 
the region right above the $\nu_e$ surface.
As one moves from the disk center towards
the edge, the MS preserving unstable solution 
shifts from one branch to another (see Figs.~\ref{fig-wkz} and \ref{fig-insdisk}),
and there is a small region 
where no MS preserving unstable solution exists. 
Nevertheless, the MS breaking unstable solution 
is nonzero in this region. 
We conclude that fast conversions can occur in 
most of the region above the merger remnant of the NS 
disks for any  realistic $\alpha$.

For the remnant system consisting of a central BH and an accretion disk,
we show in Fig.~\ref{fig-insz0} the corresponding
temporal instability with thin dashed curves for 
$R_0=0.15~R_{\nu_e}$
for both $\nu_e$ and $\bar\nu_e$
emitting tori (see Fig.~\ref{fig-gv1} and Appendix A for details). 
The growth rates are different in the proximity of 
$R_0$, but for $x\gtrsim 0.4 R_{\nu_e}$,
they coincide with the values in the
NS--disk ($R_0=0$) cases. Note that close to $R_0$, 
the growth rates might be even enhanced due to the suppressed
$\bar\nu_e$ phase space in our simple toy model.

\section{Spatial instabilities above the remnants}  \label{sec:spatial}
Similarly to the temporal instabilities discussed in Sec.~\ref{sec:temporal}, 
the ELN crossings can also lead to the occurrence of spatial 
instabilities~\citep{Dasgupta:2016dbv,Izaguirre:2016gsx}.
However, differently from the temporal instabilities whose growth rate
does not depend on the adopted matter density profile, 
the matter density has strong impact on the occurrence of 
spatial instabilities~\citep{Chakraborty:2016lct,Dasgupta:2016dbv}. 

To study  spatial instabilities in the compact binary merger
remnants, we adopt the same NS--disk neutrino emission model as in Sec.~\ref{sec:temporal}
and the cylindrically-averaged electron number density profile $n_e(x,z)$
from Ref.~\citep{Perego:2014fma} at 60~ms post merger. As we will show later, spatial instabilities
are more likely to occur in the low-density polar region above the disk, therefore we choose the location 
$(x,z)=(0,0.25R_{\nu_e})$ on the $\nu_e$ emitting surface as a benchmark example to study the DR for the occurrence of
 spatial instabilities.
\begin{figure}[t]
  \includegraphics[angle=0,width=0.85\columnwidth]{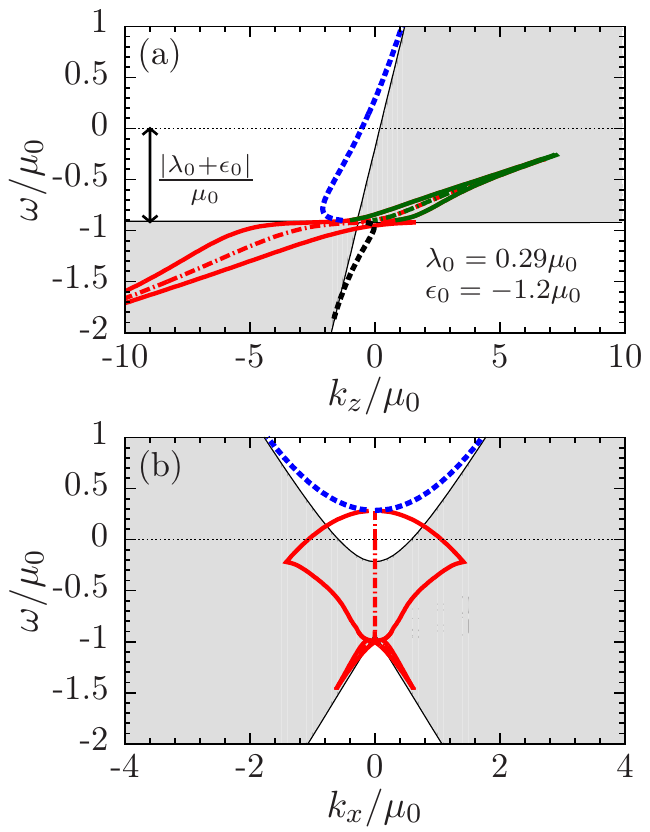}     
  \caption{Dispersion relation of $(\omega,\mathbf{k})=(\omega,0,0,k_z)$ 
  [panel(a)] and $(\omega,k_x,0,0)$ [panel(b)]
  for the ELN distribution at $(x,z)=(0,0.25R_{\nu_e})$ with
  $\alpha=\Phi^0_{\bar\nu_e}/\Phi^0_{\nu_e}=2.4$ (see top left panel of Fig.~\ref{fig-gvfull}). 
  For the complex $k_{z,x}$ that lead to the spatial flavor instability, 
  ${\rm Re}(k_{z,x})$ are shown by the red (green) dash-dotted curves
  and ${\rm Re}(k_{z,x})\pm{\rm Im}(k_{z,x})$ are shown by
  the red (green) solid curves for the MS
  preserving (breaking) solutions. The gray regions are zone of 
  avoidance for real $(\omega,k_{z,x})$.
  \label{fig-kw}}
\end{figure}
Figure~\ref{fig-kw} shows the $k_z$ ($k_x$) solution of DR 
as a function of $\omega$ for a mode with $k_x=k_y=0$ ($k_z=k_y=0$)
for the NS--disk system with $\alpha=2.4$.

Since the spatially unstable solutions usually exist at $|\omega-\lambda_0-\epsilon_0|\lesssim\mu_0$,
a large $\lambda_0+\epsilon_0$ can suppress the instability for
a mode with $\omega = 0$ as shown in the panel (a) of Fig.~\ref{fig-kw},
in which the mode propagating at the $z$ direction with $\omega=k_x=k_y=0$ is stable. 
(This is mainly due to $\epsilon_0$, similarly to the multiangle self-suppression~\citep{Duan:2010bf,Banerjee:2011fj}.)

However, this depends on the propagating mode that one is examining. 
For example, panel (b) shows that, at the same location, a spatial instability can exist 
for the mode propagating in the $x$ direction with $\omega=k_z=k_y=0$. 
This is due to the fact that neutrinos travel in all 
$\phi$ directions above the disk, therefore providing enough
transverse flux for the spatial instabilities to occur~\footnote{Similar spatial instabilities
should be expected to happen also within the SN case even when neutrinos are not propagating radially backwards.}.
Moreover, even though the propagating mode with $\omega = 0$
can be stable with respect to spatial instabilities in some cases, if there exists a highly oscillatory perturbation
in the mode with $\omega\approx \lambda_0 + \epsilon_0$, the system can always
be unstable due to the ELN crossing (see previous section).

\begin{figure}[t]
  \includegraphics[angle=0,width=1.\columnwidth]{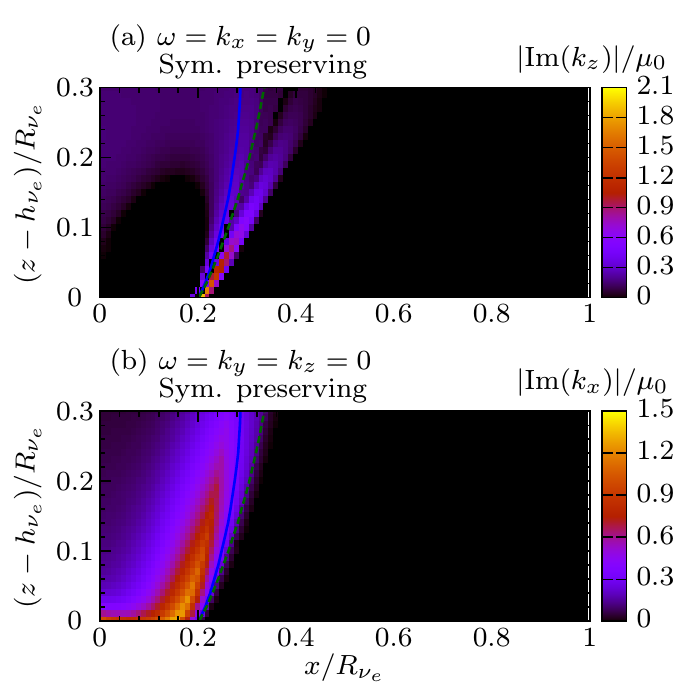}     
  \caption{Contour plot of the mirror symmetry preserving solution
  of $|{\rm Im}(k_z)|/\mu_0$ [panel(a)]
  and $|{\rm Im}(k_x)|/\mu_0$ [panel(b)] in the $(x,z)$ 
  plane above the $\nu_e$-surface for $\omega=k_x=k_y=0$ 
  and $\omega=k_z=k_y=0$ for the NS-disk model with $\alpha=2.4$,
  using the $n_e$ profiles of~\cite{Perego:2014fma} at 60~ms
  post merger. 
  Also shown are the MNR locations for neutrinos emitted at $(x,z)=(0,h_{\nu_e})$
  (solid blue curve) and at $(x,z)=(-R_{\nu_e},h_{\nu_e})$
  (dashed green curve). 
  The spatial instabilities exist inside the low-density 
  funnel where $\lambda_0\lesssim|\epsilon_0|$ and are largely suppressed
  outside the funnel where $\lambda_0\gg|\epsilon_0|$.}
  \label{fig-insdisk_spa}
\end{figure}

Now let us look at how the spatial instabilities depend on the
location above the disk.
Above the NS--disk, the angular momentum of the system keeps the 
density in the region close to the $z$ axis much lower compared to the
off-axis parts (see, e.g., Fig.~1 of Ref.~\citep{Frensel:2016fge}). 
Therefore, the occurrence of spatial instabilities 
sensitively depends on the location above the disk.
Figure~\ref{fig-insdisk_spa} shows the $y$-MS preserving instabilities
of the propagating modes with $\omega=k_x=k_y=0$ and $\omega=k_z=k_y=0$
above the $\nu_e$-surface. 

In the outer region above the disk, 
spatial instabilities are suppressed by the large matter potential, 
$\lambda_0\gg|\epsilon_0|$. Inside the low-density funnel, where
$\lambda_0\lesssim|\epsilon_0|$, the $k_z$ instability only exists
 where $\lambda_0\sim\epsilon_0$, close to the MNR region.  However, the propagating $k_x$ modes can be
unstable everywhere inside the low-density funnel. 

Because the instabilities only occur in the inner region above the
disk, the breaking of the local MS symmetry is minor. 
Therefore, the unstable $k_y$ solutions and the MS breaking $k_x$ solutions
are very similar to the $k_x$ MS preserving solutions and therefore are not discussed here.

We note that the growth rate of the spatial instabilities is
generally larger than the temporal instabilities discussed in Sec.~\ref{sec:temporal}.
This suggests that in order to fully grasp how flavor conversion develops
above the disk in the nonlinear regime, numerical simulations that 
evolve simultaneously in time and space coordinates are necessary.

\section{Discussion} \label{sec:discussion}
Within our simplified two-neutrino-emission disks model, we have shown that both  temporal and  spatial 
instabilities can
exist above the remnants of compact binary mergers.
This is a direct consequence of the larger local antineutrino 
emissivity compared to the neutrino one as well as of the spatial separation of their
respective emission surfaces.

In a more realistic environment,  the matter density profile along
the $z$ direction would decrease much faster on the surface of the central massive
neutron star and in the inner part of the disk. As a consequence,  
the separation between  $\nu_e$ and $\bar\nu_e$ in the inner disk region would  be
smaller than in the outer region. 
However, one should still expect an overall more extended $\nu_e$ surface 
with respect to the $\bar\nu_e$ one in the 
outer region 
as a result of the spatial extension of the accretion disk.

\begin{figure}[t]
  \includegraphics[angle=0,width=1.\columnwidth]{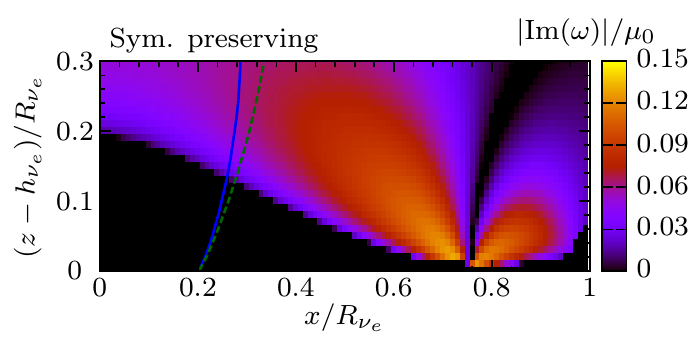}     
  \caption{Symmetry preserving solution of $|{\rm Im}(\omega)|/\mu_0$ in the $(x,z)$ 
  plane above the $\nu_e$-surface for $\mathbf{k}=0$ 
  for the same NS-disk model as in Fig.~\ref{fig-insdisk} except for
  that $h_{\nu_e}=h_{\bar\nu_e}=0.25R_{\nu_e}$. The temporal instability 
in the central part right above the disk is suppressed, see  
 Fig.~\ref{fig-insdisk} for comparison.
  \label{fig-insdisk_eqh}}
\end{figure}

In order to investigate whether the above effect could modify our main findings, we have 
examined the occurrence of instabilities in an extreme scenario 
in which the heights of $\nu_e$ and $\bar\nu_e$ 
emission surfaces are the same.
As an example, Fig.~\ref{fig-insdisk_eqh} shows the
$y$-MS preserving solution of $|{\rm Im}(\omega)|/\mu_0$ in the $(x,z)$ 
plane above the $\nu_e$ surface for $\mathbf{k}=0$
for a NS--disk model with $h_{\nu_e}=h_{\bar\nu_e}=0.25R_{\nu_e}$,
$R_{\bar\nu_e}=0.75R_{\nu_e}$, and $\alpha=2.4$.
In this  case, the temporal instability 
in the central part right above the disk can be suppressed 
(see Fig.~\ref{fig-insdisk} with $h_{\bar\nu_e}=0.75h_{\nu_e}$ for comparison).
This is because of the reduced $\nu_e$ dominated part of the ELN distribution  
very close to the inner part of $\nu_e$ emission surface.

Far away from the surface in Fig.~\ref{fig-insdisk}, where the
spatial extension of the disks becomes important, favorable  conditions
for temporal instability are however recovered.
Furthermore, the spatial instabilities for the transversely propagating
mode can always exist.
Thus, we conclude that even taking into account a more realistic 
neutrino emission geometry, flavor instabilities leading to
fast pairwise conversion exist in an extended region above the merger remnants.

In a realistic astrophysical environment, a non-negligible flux of
neutrinos streaming in the negative $z$ direction exists because of collisions.
This may also be responsible for fast 
flavor conversions~\cite{Chakraborty:2016lct,Dasgupta:2016dbv,Izaguirre:2016gsx}.
Given the still preliminary neutrino transport employed in 
merger simulations with respect to SN simulations, we refrain from including any 
backward contribution in our toy-model. 
In fact, given the favorable conditions for fast flavor conversions 
determined by the ubiquitous crossings of ELN distribution, 
any contribution due to neutrino collisions 
would further strengthen our conclusions. 

Likewise, hydrodynamical simulations would predict location-dependent neutrino emissivities, 
forward-peaked neutrino angular distributions on their emission
surfaces (instead of the uniform distributions adopted here), and time-dependent neutrino 
properties due to the long-term evolution of the remnant system.
Those effects, not considered in our simple model, can all
quantitatively influence the exact conditions responsible for
fast conversions. However, we expect the 
qualitative features presented here will be unchanged.

%%%%%%%%%%%%%%%%%%%%%%%%%%%%%%%%%%%%%%%%%%%%%%%%%%%%%%%%%%%%%%%%%%%%%%%%%%%%%%%
% Conclusions
%%%%%%%%%%%%%%%%%%%%%%%%%%%%%%%%%%%%%%%%%%%%%%%%%%%%%%%%%%%%%%%%%%%%%%%%%%%%%%%

\section{Conclusions}\label{sec:conclusions}
NS--NS and NS--BH compact mergers are neutrino-dense 
sources. For the first time, we investigate the role of the neutrino 
angular distributions in flavor conversions in merger 
remnant disks and explore the occurrence of fast flavor conversions. 
Given the current uncertainties on the modeling of the 
neutrino transport in merger remnants,
we do not rely on direct inputs from a specific hydrodynamical 
simulation and, instead, model the local neutrino radiation field 
using a simple two-neutrino-emitting surfaces model 
whose parameters are guided by \citep{Perego:2014fma}.
We then adjust the model 
parameters to cover the range spanned by existing 
hydrodynamical simulations and obtain reliable results. 

Because of the larger $\bar{\nu}_e$ flux with respect to the $\nu_e$ one, 
and the emission geometry of the disk, 
crossings between the 
$\nu_e$ and $\bar{\nu}_e$ angular distributions are present 
for any point above the merger disk remnant, differently from SNe,
where crossings do not occur unless 
strong directional-dependent neutrino emission is invoked due to the 
multidimensional effects. 
Consequently, we showed in Sec.~\ref{sec:temporal} that
fast flavor conversions caused by temporal disturbances
may be intrinsically unavoidable in
compact mergers remnants.

We also discussed the presence of spatial instabilities in Sec.~\ref{sec:spatial}. 
Those are sensitive to the model-dependent matter density profile and 
to the  transversally propagating neutrinos. 
Favorable conditions for those instabilities, with growth rates larger than the temporal 
instabilities, appear above merger remnants.
Nevertheless, spatial instabilities occur in 
smaller spatial regions than the temporal ones due to the partial matter suppression effect 
(unless initial perturbations exists in $\omega \approx \lambda_0 + \epsilon_0$  not studied in this work).
The presence of any spatial instabilities would make 
the system, already suffering unavoidable temporal instabilities, even more unstable.
Hence, it is necessity to  take
into account both temporal and spatial evolution of flavor conversion 
in upcoming numerical studies.

A more realistic shape of neutrino emitting surfaces may
certainly affect the results obtained within our simple two-disks model quantitatively.  
In Sec.~\ref{sec:discussion}, we discussed the potential impact and showed that 
our conclusions should be qualitatively fairly general
despite that the two-disk neutrino emission model is highly simplified.

The stability analysis presented in this paper proves 
that favorable conditions for fast flavor 
conversions can exist in mergers. However, similarly to 
the case of core-collapse supernovae, 
we still miss an exact numerical solution of the neutrino 
flavor evolution. Our preliminary results suggest that fast flavor conversions should 
be taken seriously in compact merger remnants as 
they may lead to flavor equilibration very close to 
the neutrino decoupling region.

Flavor equilibration in the neutrino decoupling region could be responsible for nontrivial 
modifications of the jet dynamics, if powered by neutrinos,
as well as of the production of heavy elements. For example, it could
largely reduce the neutrino pair annihilation rate above a 
BH-torus remnant~\cite{Ruffert:1996by,Foucart:2015vpa}.
Similarly, it could also change the composition of
the ejecta exposed to neutrinos~\cite{Wanajo:2014wha,Metzger:2014ila,Perego:2014fma}.
This will modify the predicted $r$-process outcome and the associated
kilonova light curves in merger events.
Given that the relevant length scales induced 
by fast conversions are much smaller than the resolution of merger 
simulations, the impact of pairwise conversions should be phenomenologically explored. 
In addition, as a consequence of our findings, the  
adopted picture of the MNR occurring in merger remnants should 
be revised, and future work is needed in this direction.

%%%%%%%%%%%%%%%%%%%%%%%%%%%%%%%%%%%%%%%%%%%%%%%%%%%%%%%%%%%%%%%%%%%%%%%%%%%%%%%
%Acknowledgements %%%%%%%%%%%%%%%%%%%%%%%%%%%%%%%%%%%%%%%%%%%%%%%%%%%%%%%%%%%%%
%%%%%%%%%%%%%%%%%%%%%%%%%%%%%%%%%%%%%%%%%%%%%%%%%%%%%%%%%%%%%%%%%%%%%%%%%%%%%%%
\acknowledgments

We thank Thomas Janka for helpful discussions, Albino Perego for 
providing the density profiles of the simulation presented in 
Ref.~\cite{Perego:2014fma}, as well as Ignacio Izaguirre 
and Georg Raffelt for useful comments on the manuscript. 
Support from the Knud H{\o}jgaard Foundation, the Villum Foundation (Project
No.\ 13164) and the Danish National Research Foundation (DNRF91) is acknowledged.

\begin{appendix}

%\newpage
%\clearpage

%\onecolumngrid

\twocolumngrid

\section{Emission geometry of the two-disks neutrino model}
The toy-model adopted to simulate the neutrino 
emission from compact binary merger remnants is here described in detail. 
The local neutrino radiation field is 
also characterized for any point above the 
emitting surface. 

The neutrino decoupling region has been approximated
by flavor-dependent emitting surfaces 
shown in Fig.~\ref{fig-gv1}. 
For the sake of simplicity, 
we assume that the (energy-integrated) angular distribution 
on each $\nu$ surface is half-isotropic with 
$\Phi_{\nu_\beta}^0=dn_{\nu_\beta}^0/d\Omega=
L_{\nu_\beta}/[2\pi^2(R_{\nu_\beta}^2-R_0^2)\langle E_{\nu_\beta}\rangle]$
for $\cos\theta\geq 0$, $R_0\leq\sqrt{x^2+y^2}\leq R_{\nu_\beta}$ at $z=h_{\nu_\beta}$,
where $L_{\nu_\beta}$ is the total energy luminosity of $\nu_\beta$ and 
$\langle E_{\nu_\beta}\rangle$ is the corresponding average energy. 
The choice of the model parameters is guided by 
the hydrodynamical simulation of the massive NS--disk
evolution as a remnant of a NS--NS merger~\cite{Perego:2014fma}:  
$R_{\bar\nu_e}=0.75R_{\nu_e}$,
$h_{\nu_e}/R_{\nu_e}=h_{\bar\nu_e}/R_{\bar\nu_e}=0.25$, and
$\alpha\equiv\Phi^0_{\bar\nu_e}/\Phi^0_{\nu_e}=2.4$.

At a given location, $P \equiv (x,z)$, above the neutrino emitting surfaces depicted 
in Fig.~\ref{fig-gv1}, $g(\mathbf{v})$ can be rewritten as
\begin{equation}\label{eq-gdisk}
g(\mathbf{v})=\sqrt{2}G_F[\Phi^0_{\nu_e}\zeta_{\nu_e}(\mathbf{v})-
\Phi^0_{\bar\nu_e}\zeta(\mathbf{v})]\ ,
\end{equation}
where $\zeta_{\nu_\beta}(\mathbf{v})=1$
for any $\mathbf{v}$ that can be traced back to the corresponding 
$\nu_\beta$-emitting surfaces and 0 otherwise. 

\begin{figure}[t]
  \includegraphics[angle=0,width=0.9\columnwidth]{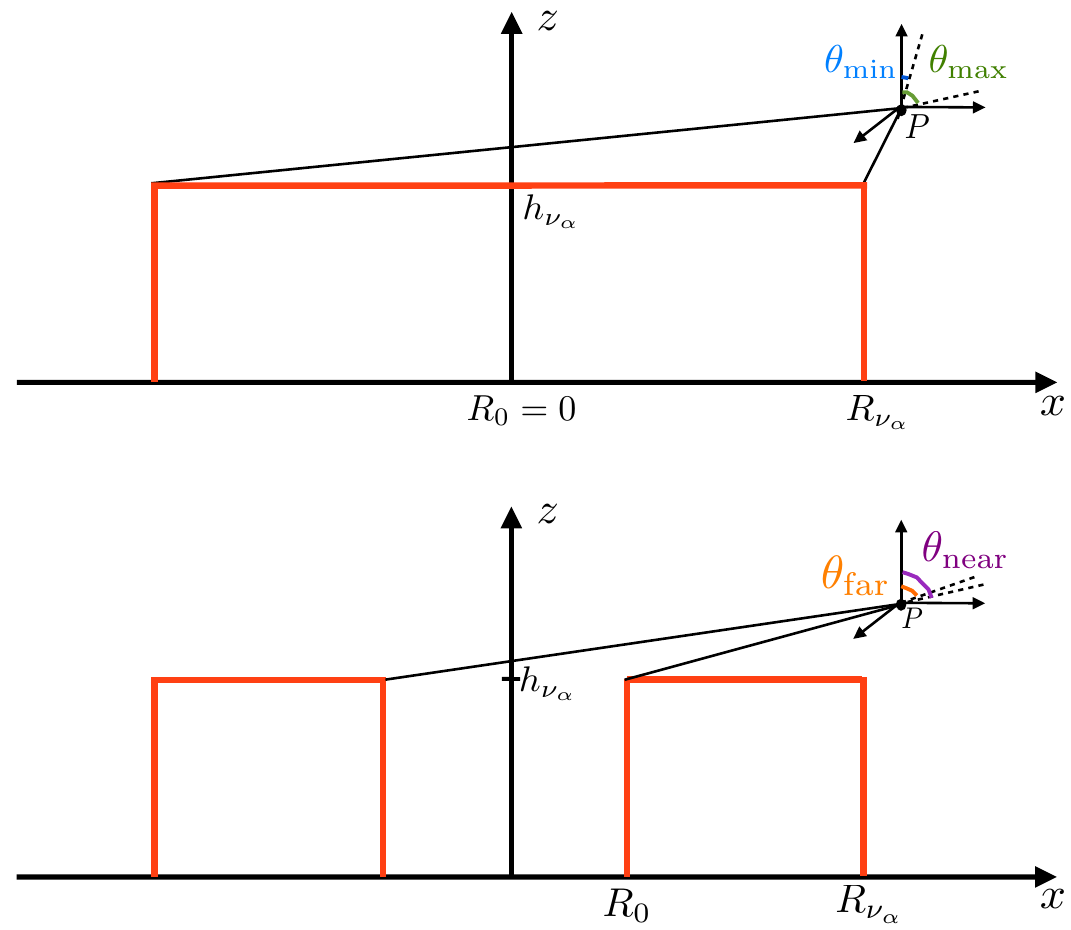} \\
   \includegraphics[angle=0,width=0.9\columnwidth]{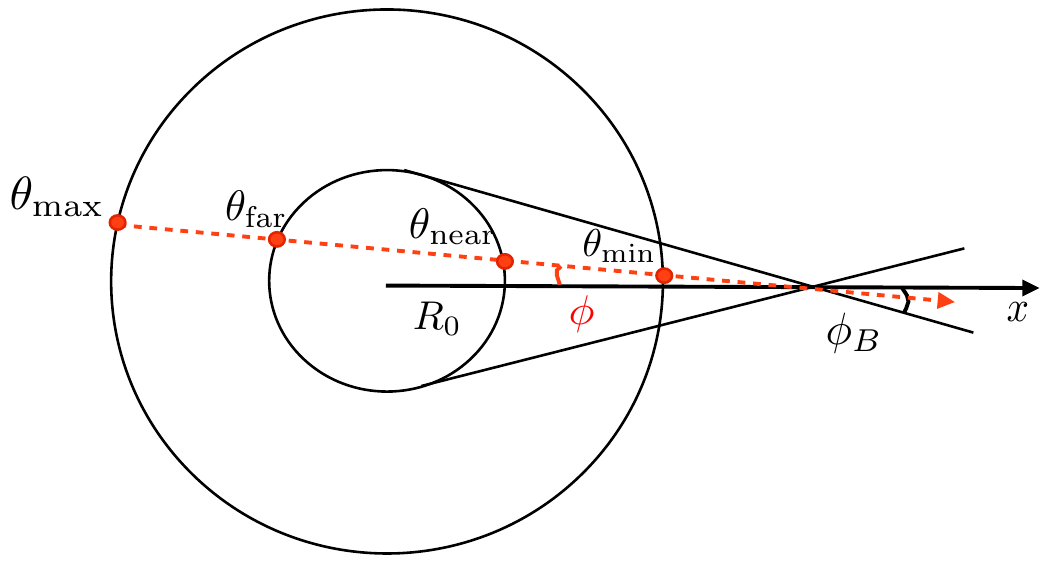}      
  \caption{Angular projections for neutrinos %the neutrino distributions 
  emitted from the disk surface. 
  Top panel:  
  The neutrino distribution at a 
  point $P$ at a certain distance from the NS-disk surface 
  is restricted within the angles $\theta_{\rm min}$ and $\theta_{\rm max}$
  for a certain $\phi$ [see Eq.~\eqref{eq:thetaminmanx}]. 
  Middle panel: 
  Same as above but for a BH-disk surface. 
  The neutrino distribution at $P$ also depend 
  on $\phi$. In this example, neutrinos can only reach $P$
  from $\theta_{\rm min}\leq\theta\leq\theta_{\rm near}$ 
  and $\theta_{\rm far}\leq\theta\leq\theta_{\rm max}$,
  see Eq.~\eqref{eq:thetanearfar}. 
  Bottom panel: Angular projections for the BH-disk viewed from above, 
  see text for details.  
  \label{fig:thetaangles}}
\end{figure}

For the NS--disk  $(R_0=0)$,
the ranges of $\theta\in [\theta_{\rm min},\theta_{\rm max}]$ 
and $\phi \in [-\phi_{\rm max},\phi_{\rm max}]$ in which 
$\zeta_{\nu_\beta}(\mathbf v)=1$ can be derived straightforwardly
and are shown in Fig.~\ref{fig:thetaangles} (top panel).
They are defined by
\begin{subequations}
\begin{align}
&\cos\phi_{\rm max}=
\begin{cases}
-1 & 0\leq x<R_{\nu_\beta}\\
\dfrac{\sqrt{x^2-R_{\nu_\beta}^2}}{x} & x\geq R_{\nu_\beta}\ ,
\end{cases}\\
&\tan{\theta_{\rm min}}=
\begin{cases}
0 & 0\leq x<R_{\nu_\beta}\ \\
\dfrac{x\cos\phi-\sqrt{R_{\nu_\beta}^2-x^2\sin^2\phi}}{z-h_{\nu_\beta}} 
& x\geq R_{\nu_\beta}\ ,
\end{cases}\\
& \tan\theta_{\rm max}=\frac{x\cos\phi+\sqrt{R_{\nu_\beta}^2-x^2\sin^2\phi}}{z-h_{\nu_\beta}}\ .
\end{align}
\label{eq:thetaminmanx}
\end{subequations}
Note that $\theta_{\rm min}$ and $\theta_{\rm max}$ are
functions of $\phi$.

For the BH--torus, the corresponding values of $\phi_{\rm max}$ 
are the same as in the above equations. 
On the other hand, the range of $\theta$ for $x\geq R_0$ becomes more complicated. 
Given the configuration of the emitting surface shown in  Fig.~\ref{fig:thetaangles} (middle panel),
for $\phi \geq\phi_B = \arccos(\sqrt{x^2-R_0^2}/x)$ or $\phi\leq -\phi_B$,
$\zeta_{\nu_\beta}(\mathbf v)=1$ for
$\theta\in [\theta_{\rm min},\theta_{\rm max}]$.
Similarly, for $-\phi_B<\phi<\phi_B$, $\zeta_{\nu_\beta}(\mathbf v)=1$ for
$\theta\in [\theta_{\rm min},\theta_{\rm near}]$ and 
$\theta\in [\theta_{\rm far},\theta_{\rm max}]$ with 
\begin{subequations}
\begin{align}
&\tan{\theta_{\rm near}}=
\frac{x\cos\phi-\sqrt{R_0^2-x^2\sin^2\phi}}{z-h_{\nu_\beta}}\ ,\\
&\tan{\theta_{\rm far}}=
\frac{x\cos\phi+\sqrt{R_0^2-x^2\sin^2\phi}}{z-h_{\nu_\beta}}\ .
\end{align}
\label{eq:thetanearfar}
\end{subequations}
See also the bottom panel of Fig.~\ref{fig:thetaangles} for details. 

\begin{figure*}[t]
  \includegraphics[angle=0,width=2\columnwidth]{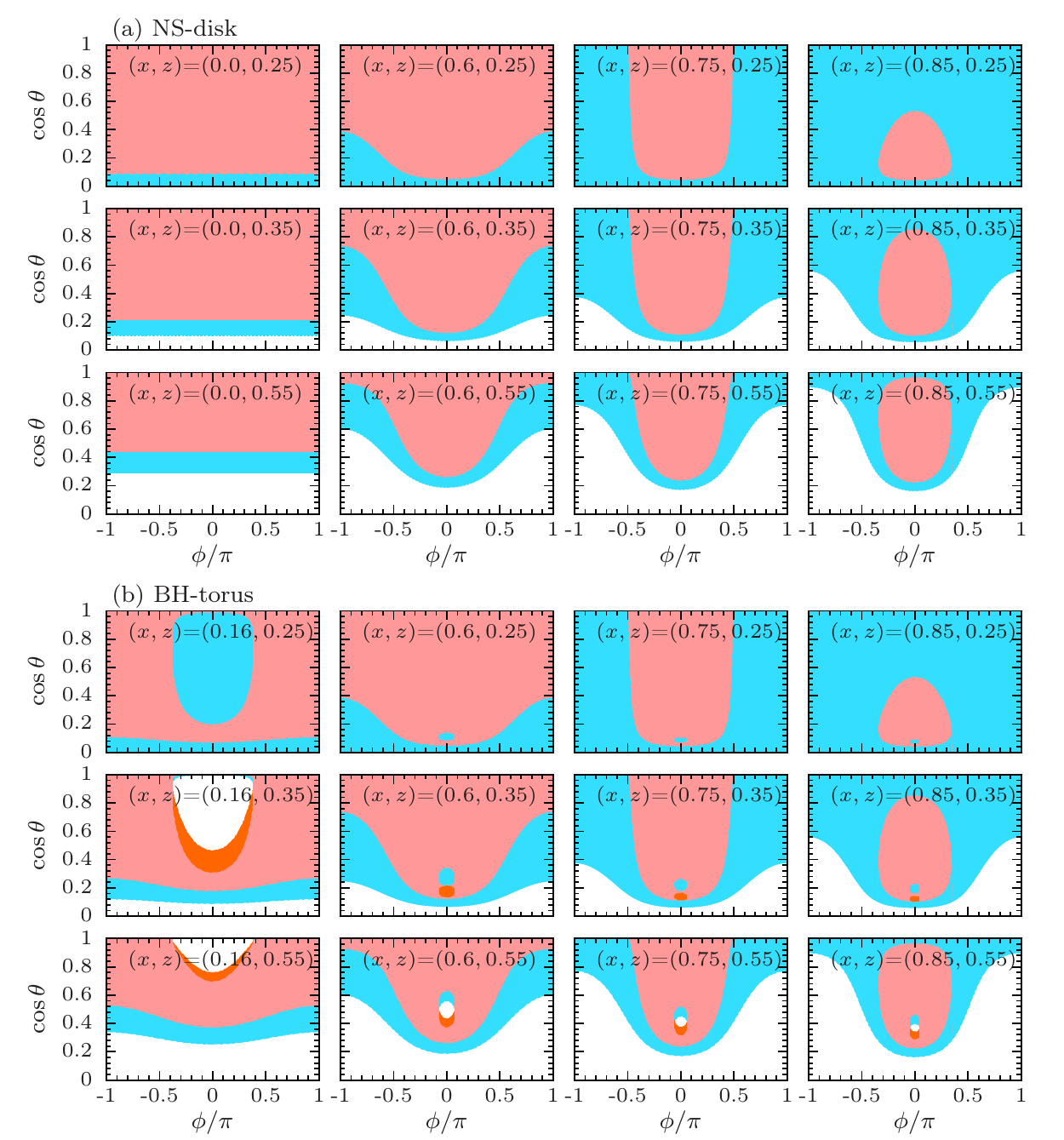}     
  \caption{Electron neutrino lepton number distribution, 
  $\Phi_{\nu_e}-\Phi_{\bar\nu_e}$, as a function of
  $\cos \theta$ and $\phi$ and different $(x,z)$ points above 
  the $\nu_e$ surface. The top three rows [panel (a)] refer to 
  our benchmark NS--disk model ($R_{\bar\nu_e}=0.75R_{\nu_e}$ 
  $h_{\nu_e}/R_{\nu_e}=h_{\bar\nu_e}/R_{\bar\nu_e}=0.25$, and 
  $\alpha\equiv\Phi^0_{\bar\nu_e}/\Phi^0_{\nu_e}>1$). 
  In the red (blue) shaded area, $\Phi_{\nu_e}-\Phi_{\bar\nu_e}<0$ 
  ($\Phi_{\nu_e}-\Phi_{\bar\nu_e}>0$). 
  Null electron neutrino lepton number is shown in white. 
  The bottom three rows [panel (b)] refer to the BH--torus case 
  with $R_0=0.15 R_{\nu_e}$ and the orange shaded area marks 
  regions where $\Phi_{\nu_e} = 0$. In both cases, as one moves away from 
  the surface center, the $\bar\nu_e$ contribution is significantly reduced.
  \label{fig-gvfull}}
\end{figure*}

As an illustrative example, 
$\Phi_{\nu_e}-\Phi_{\bar\nu_e}= [\Phi^0_{\nu_e}\zeta_{\nu_e}(\mathbf{v})-
\Phi^0_{\bar\nu_e}\zeta(\mathbf{v})]$ 
as a function of $\cos \theta$ and $\phi$ is shown in Fig.~\ref{fig-gvfull}  
for the same NS--disk model of Fig.~\ref{fig-gv1} (panel (a), top three rows) 
and for the BH--torus model (panel (b), bottom three rows).  
A set of $(x,z)$ locations is plotted for both the NS and BH cases. 
In the red shaded area, $g(\mathbf{v})<0$ 
[$\zeta_{\nu_e}(\mathbf v)=\zeta_{\bar\nu_e}(\mathbf v)=1$] 
for $\Phi^0_{\bar\nu_e}/\Phi^0_{\nu_e} > 1$. 
In the blue area, $g(\mathbf{v})>0$ [$\zeta_{\bar\nu_e}(\mathbf v)=0$] .
In the white area, $g(\mathbf{v}) = 0$ and in the 
orange area $\zeta_{\nu_e}(\mathbf v)=0$. 
Figure~\ref{fig-gvfull} shows that as one moves away from the surface center,
the $\bar\nu_e$ contribution can be significantly reduced and $g(\mathbf{v})=0$ 
for a certain range of $\cos\theta$ and $\phi$. 
In the BH--torus  ($R_0\neq 0$), white holes appear in $g(\mathbf{v})$ as well as
regions only populated by $\bar{\nu}_e$ because of 
the toroidal geometry described above. 

\end{appendix}

\bibliography{insdisk}

\end{document}